# Bond Alternation, Polarizability and Resonance Detuning in Methine Dyes

Seth Olsen[*] and Ross H. McKenzie

*School of Mathematics and Physics and Centre for Organic Photonics and Electronics, The University of Queensland, Brisbane, QLD 4072 Australia*

*Email: s.olsen1@uq.edu.au

**Keywords:** *Structure-property relationships, Brooker dyes, methine dyes, push-pull polyenes, nonlinear optical response, bond length alternation, bond order – bond length relationships, fluorescent proteins*

*Abstract*

We derive structure-property relationships for methine ("Brooker") dyes relating the color of the dye and its symmetric parents to its bond alternation in the ground state and also to the dipole properties associated with its low-lying charge-resonance (or charge-transfer) transition. We calibrate and test these relationships on an array of different protonation states of the green fluorescent protein chromophore motif (an asymmetric halochromic methine dye) and its symmetric parent dyes. The relationships rely on the assumption that the diabatic states that define the Platt model for methine dye color [J.R. Platt, *J. Chem. Phys.* **25** 80 (1956)] can also be distinguished by their single-double bond alternation and by their charge localization character. These assumptions are independent of the primary constraint that defines the diabatic states in the Platt model – specifically, the Brooker deviation rule for methine dyes [L.G.S. Brooker, *Rev. Mod. Phys.* **14** 275 (1942)]. Taking these assumptions, we show that the Platt model offers an alternate route to known structure-property relationships between the bond length alternation and the quadratic nonlinear polarizability $\beta$. We show also that the Platt model can be parameterized without the need for synthesis of the symmetric parents of a given dye, using dipole data obtained through spectroscopic measurements. This suggests that the Platt model parameters may be used as independent variables in free-energy relationships for chromophores whose symmetric parents cannot be synthesized *or chromophores strongly bound to biomolecular environments*. The latter category includes several recently characterized biomolecular probe constructs. We illustrate these concepts by an analysis of previously reported electroabsorption and second-harmonic generation experiments on green fluorescent proteins.

**I. Introduction**

Molecules with a high non-linear optical response are technologically useful,[1] and it is worthwhile to understand relationships between molecular structure and optical response in these systems.[1,2] There have been recent interesting advancements in the application of nonlinear chromophores to problems in biological imaging.[3] Imaging by excitation of two-photon fluorescence and second-harmonic generation is useful because the intensity-dependence of these phenomena allow localized excitation areas and because light at longer wavelengths can more easily penetrate biological tissue.[4]

The first and second order nonlinear optical response coefficients $\alpha$ and $\beta$ determine the facility of two-photon absorption and second-harmonic generation, respectively.[5] The discovery of new applications of these phenomena has driven interest in new structure-property relationships for characterizing the optical response. In particular, it has been shown that the nonlinear optical responses of conjugated molecules can be related to descriptors based on the single-double bond alternation and molecular dipole properties.[6,7-9]

There have been recent significant advances in the development of specialized chromophores and dyes for biophotonic imaging techniques. Of particular interest are dyes whose optical properties (particularly, fluorescence quantum yield) are modified by binding to biomolecular environments, as these chromophores offer the ability to selectively image specific binding targets.[10]

The molecular structures of fluorogenic dyes are related to those of dyes with high optical nonlinearities, as members of both classes contain heterocyclic rings separated by an unsaturated carbon bridge.[11,12] Dyes containing this structural motif are sometimes called "Brooker dyes" after L.G.S. Brooker, whose work enhanced our early understanding of color-structure relationships in these systems.[13,14] By the mid-twentieth century, there was a considerable body of established results relating to structure-property (particularly, structure-color) relationships in these systems.[15-17] Olsen has recently shown that there is a family of solutions to the state-averaged complete active space self-consistent field problem for dyes in this class, which duplicates the Brooker-Platt color-constitution relationships, with respect to both its conceptual structure and its quantitative predictions.[18]

The purpose of this article is to show that key concepts from the color-constitution relationships for Brooker dyes can also be used to formulate structure-property relationships for the non-linear optical response. These relationships connect the well-established relationships characterizing Brooker dye color[15,16] with more recent work[2,6,7-9,19,20] in structure-property relationships for non-linear chromophores.

We will examine a simplified parametric ("essential state"[21]) model proposed by Platt[16] for the properties of Brooker dyes, and compare these with *ab initio* quantum chemical calculations. Parametric models and *ab initio* quantum chemistry are complementary approaches to electronic structure. The former's strength is the ability to highlight trends that span families of related molecules (represented on different Hilbert spaces), while the strength of the latter is an objective assessment of the character of the electronic (Born-Oppenheimer) eigenstates for a particular case. *Ab initio* approaches can be used as a check on essential state models by identifying when states with the required character do not fall in a low-energy subspace. Essential state models can identify when *ab initio* approximations fail by identifying calculations that lie outside the bounds of well-characterized reaction series. Both approaches are indispensable in the pursuit of a coherent chemical picture.

The paper will proceed as follows. In section II, we will introduce a two-state model of absorption in methine dyes ("Brooker dyes") proposed by Platt,[16] and describe two concepts which are central to the model: the *Brooker basicity difference* and the *isoexcitation energy*.[15] In section III we derive relationships for the bond order and bond length alternation in methine dyes that depend on the Brooker basicity difference. In section IV we will do the same for dipole properties (specifically, the transition and difference dipole observables). In section V we discuss relationships for the first and second order optical responses, which emerge when the bond length and dipole property formulas derived from Platt's model are incorporated into more recent structure-property relationships which describe these quantities.[2,5,22] Sections III-IV will contain illustrative comparisons against a quantum chemical data set, which is described in detail in Section VI. Sections VII and VIII provide a discussion and summary conclusion.

## II. Platt's Two-State Model of Color in Brooker Dyes

"Brooker dyes" are methine (cyanine-like) dyes, whose common structural motif is two heterocyclic "nuclei"[23] separated by an unsaturated carbon chain bridge. Brooker described an empirical rule for the color of such dyes, wherein the maximum absorbance wavelength of an asymmetrical dye (having different nuclei at the ends) is no redder than the mean absorbance of the "parent" symmetric dyes, each of which possesses two copies of one of the nuclei present in the asymmetric dye.[15,16] The energy corresponding to the mean wavelength is the harmonic mean of the excitation energies of the parents, and is called the "isoexcitation energy", $E_I$. The *Brooker deviation rule* states that the excitation energy $E_{LR}$ of the asymmetric dye with nuclei L and R obeys

$$E_{LR} \geq E_I \equiv 2\left(\frac{1}{E_{LL}} + \frac{1}{E_{RR}}\right)^{-1} \tag{1}$$

where $E_{LL}$ and $E_{RR}$ are the excitation energies of the symmetric parent dyes carrying nuclei L and R respectively. These concepts are illustrated in Figure 1.

In general, the excitation energy of the asymmetric dye will be greater than $E_I$. Brooker noted that the color of asymmetric dyes could be correlated with the kinetics of the condensation reaction which formed them.[24] He concluded deviation from the equality measures a chemical energy associated with the addition or removal of an electron from the heterocycle. Brooker formulated several "basicity" scales by ordering the nuclei by measuring the deviation from the isoexcitation wavelength ("Brooker Deviation") in different dyes.[15] The Brooker Deviation can be correlated with other measures of charge accepting/donating ability, such as the Hammet $\sigma_m$ parameter.[13]

Platt considered a two state empirical Hamiltonian model of the excitations of Brooker dyes.[16] Platt's Hamiltonian is written

$$H^{Platt} = \frac{1}{2}\begin{pmatrix} b_{LR} & E_I \\ E_I & -b_{LR} \end{pmatrix} \quad (2)$$

where $b_{LR}$ is the "*Brooker basicity difference*" characterizing the nuclei of the dye, and is defined as

$$b_{LR} \equiv \pm\sqrt{E_{LR}^2 - E_I^2} \quad (3)$$

so that, by construction, the excitation energy of the asymmetric dye in question is given as the difference in eigenvalues of $H^{Platt}$. One of several important contributions made in Platt's 1956 paper was the demonstration that Brooker's spectroscopic data was consistent with writing

$$b_{LR} = b_R - b_L \quad (4)$$

so that *the Brooker basicity difference is a difference between basicities that are transferrable properties of each nucleus.*[16]

Platt's model implicitly defines two diabatic electronic basis states $|L\rangle$ and $|R\rangle$ by partitioning the diagonal and off-diagonal elements of $H^{Platt}$ according to the Brooker deviation rule. As in any two-state model, the transformation onto the eigenstate basis is characterized by a single angle $\theta^{Platt}$. This is uniquely specified by a "detuning parameter" $\lambda^{Platt}$, the ratio of the Brooker basicity difference and the isoexcitation energy.

$$\lambda^{Platt} = \cot 2\theta^{Platt} \equiv \frac{b_{LR}}{E_I} \quad (5)$$

The detuning parameter $\lambda^{Platt}$ will emerge as a key concept in developments below.

Platt's model is based on the Brooker's *empirical* results. Platt knew this, and pointed out explicitly[16] that the equations defining his model were a direct "working equation" transcription of Brooker's[15] key results. Aspects of the model have never been satisfactorily derived from first principles. In particular, the relationship specifying the isoexcitation energy as a *harmonic* mean of the parent symmetric dye excitations has never been derived as a consequence of more fundamental principles. Quantum mechanical models of Brooker dyes have usually[25] indicated an *arithmetic* mean as specifying the corresponding red limit, and have gone on to point out that the difference between these is small on the energy scales characteristic of Brooker dye absorption. On the scale of the dyes we use as examples in this paper (c.f. Section VI), the excitations of symmetric dyes are bounded by 2.29 eV and 3.09 eV. The difference between the harmonic and arithmetic means for these energies amounts to 0.06eV, which is lower than the a priori expected accuracy of the method used to calculate the energies.[26] For any set of positive numbers, the harmonic mean is always less than or equal to the arithmetic mean (with equality only if the numbers are equal), so that the harmonic mean is clearly a safer choice for expressing an empirical "red limit" rule. All the same, Brooker did test his analyses using different means and found that an arithmetic mean led to less consistent results, both with respect to the ordering of basicities of nuclei and violations of the deviation rule.[15] We use the harmonic mean here in order to maintain the connection to Brooker's work[15] and to Platt's proposed empirical model.[16]

*The Nature of Diabatic states in the Platt Model*

Techniques for defining diabatic states based upon the physically motivated constraint of expectation values of an observable have been discussed.[27] An excellent example is (Cave and Newton's) Generalized Mulliken-Hush approach, which defines the diabatic states as those which diagonalize an electronic dipole operator.[28] The diabatic states in Platt's model are different from these in two fundamental ways. Firstly, the Platt model is parameterized by *excitation* energies, not *state* energies. Constraints defining the diabatic states cannot be written in terms of the matrix elements (or eigenvalues) of any operator acting on a single state in a Hilbert space. Secondly, the excitation energies involved in the definition *are those of different dyes*, so that the relevant quantities *are calculated on different Hilbert spaces*. For this reason, the formal specification of constraints defining the diabatic states in the Platt model is difficult.[29]

The basis states in the Platt Hamiltonian are defined *only* by the fact that the associated adiabats reproduce Brooker's deviation rule. It is, however, possible to infer additional physically (and *chemically*) motivated constraints that the diabatic states should fulfil in "normal" methine dyes.[30] Firstly, if one supposes that the diabatic states correspond to different Lewis structures in a charge resonance pair (as Platt clearly did[16]), then one may suppose that each of the diabats has a definite and opposite bond order alternation. Secondly, one may also suppose that since the formal charge centres in the different Lewis structures are oppositely situated, the diabatic states should approximately diagonalize a dipole operator oriented along the vector separating the charge centres and measured with respect to a well-chosen common origin (the dyes in question are often charged). This latter point has been invoked explicitly by Simpson.[31] In this paper, we investigate the consequences that arise when these auxiliary constraints are assumed to hold for the diabatic states in the Platt model.

## III. Bond Alternation Index From Platt's Model

Platt's model Hamiltonian defines a diabatic representation for the electronic structure of a dye. The adiabatic ground state can then be written

$$|S_0\rangle = \cos\theta^{Platt}|L\rangle + \sin\theta^{Platt}|R\rangle \tag{6}$$

Assuming that the diabatic states $|L\rangle$ and $|R\rangle$ represent ideal and complementary single/double bond alternation schemes, we can define a bond order alternation parameter $x$ as the difference in population of $|L\rangle$ and $|R\rangle$.

$$x \equiv \cos^2\theta^{Platt} - \sin^2\theta^{Platt} = \cos 2\theta^{Platt} = \frac{\lambda^{Platt}}{\sqrt{1+\left(\lambda^{Platt}\right)^2}} \tag{7}$$

There are many ways to approach relationships between bond order and bond length.[32] If the alternation in bond order is small relative to a suitably chosen reference then the dependence will be linear. We can then express the length of a given bond $i$ as

$$r_i = (1 + Sign(i) * c_i x) r_{i,0} \tag{8}$$

relative to the reference state where the bond length is $r_{i,0}$. Here $c_i$ is a proportionality constant relating the *bond order alternation* to the *bond length deviation*, and Sign($i$) is ±1 depending on the mutual bond-bond polarizability[33] of bond $i$ relative to a reference bond for which the deviation is positive.

Equation 8 shows that the bond length in the reference state is that for a dye with vanishing $b_{LR}$. Clearly, this is true for a symmetric dye, where the nuclei are the same. *This identifies the appropriate reference bond length for a given nucleus L as the corresponding symmetric dye L:L.* However, we can extend this further. Equations 7 and 8 imply that the bond lengths of a given nucleus should be the same *in any resonant dye containing that nucleus.* This relationship is supported by the optimized geometries in our example data set. We show this in Figure 2 for four representative dyes: two near resonance ($\lambda^{Platt} \sim 0$) and two far from it ($\lambda^{Platt} \sim \pm 1$). This figure supports the prediction of Platt's model that the appropriate origin to use in bond length alternation studies of a methine dye is the reference state provided by the parent symmetric dyes. Although this result is quite direct and simple, we have not found previous explicit reference to it in the literature.

*We emphasize that the bond lengths in an asymmetric resonant dye are not equal, even on the bridge*. For this reason a pure bond length alternation is not an appropriate coordinate with which to measure of the deviation from resonance in methine dyes. The bond *order* alternation may still be useful, but this quantity is measured by alternation with respect to the bond lengths of the parent symmetric dyes, and not to an average bond length, nor an idealized single or double bond length. This point is clearly most important with respect to dyes with a short methine chain (such as those in our example set), because as the chain length increases, the bonds in the middle of the chain will become less dependent on the chemical identity of the distant nuclei. As the local environment of the alternant bonds becomes less distinguishable, the bond lengths should converge to a single value.

In addition to an appropriate origin for bond length deviation coordinates, we have derived in equations 7 and 8 a quantitative structure-property relationship for the dependence of the bond length on $\lambda^{Platt}$. This relationship contains one constant representing the response of the bond to the basicity difference, which can be invoked as an adjustable parameter in fitting the relationship to our data set of example dyes. Figure 3 displays these fits for four nuclei in the set: PhO-, PhOH, ImO- and ImOH. Quantitative estimates of the $c_i$ and associated errors are listed in Tables 1a and 1b.

The relationship (eqn. 8) provides a qualitatively accurate description of the bond length changes. The quantitative agreement varies. For the PhO- and PhOH nuclei it is good, providing an adequate quantitative approximation for the variation of the lengths of all bonds in the nucleus. This is not suprising, because these nuclei are derived from alternant hydrocarbon motifs via an electronegativity perturbation representing the switch of a carbon atom for an oxygen atom. Bond length/bond order relationships are considerably simplified for alternant systems.[34] On the other hand, the imidazoloxy nuclei ImO- and ImOH cannot be related to alternant hydrocarbon motifs because the ring has an odd number of sites. In this case the relationship is quantitatively much more accurate for bonds that are directly involved in the oxonol system (the chain connecting the imidazoloxy oxygen to the methine bridge). Quantitative utility deteriorates for other bonds.

The results for the ImNH nucleus (not shown in Figure 3) are not good, but the reason is easy to understand: when the imine nitrogen is protonated in this model, there is a second ancilliary resonating system created over the bonds in the N3-C4-N5 system. Although this system is disjoint from the main oxonol system, they should interact through the attraction of their respective formal charges. This coupling will change the electronic structure of the N-C-N subsystem in response to the position of the formal charge in the main oxonol chain. This has the effect of making the basicity of the imidazolinone weakly dependent on the precise state of the oxonol chain, so that the bond alternation no longer depends so strongly on the balance of the oxonol resonance alone. The relationship derived still holds qualitatively, because the interaction between the resonators is weak.

**IV. Dipole Matrix Elements from the Platt Model**

In this section we derive relationships between the color of methine dyes and dipole observables associated with the transition – specifically, the difference between adiabatic dipole expectation values and transition dipole expectation values. In order to do this, we point out (as Simpson has[31]) that the chemical structure of a cyanine or diarylmethane dye is such that the centres which bear formal charge in the resonance are often far apart. This implies that the structures should be distinguishable by a dipole operator oriented along the vector separating the centres, and measured with respect to a reasonably chosen origin (for example, the centre of charge of the nuclear frame, which will not depend on which Born-Oppenheimer electronic state is used in the estimator). The choice of origin is important because the dyes we examine can be charged.

The assumption that the resonating structures approximately diagonalize a dipole operator means that the Platt diabatic states are approximate diabatic states for a Generalized Mulliken-Hush (GMH) model. The constraint which defines the diabatic states in (Cave and Newton's) GMH model is that the transition dipole between the diabatic states vanish.[28] This represents an abstraction from earlier models after which the technique was named.[35] In a two-state model, the requirement that the transition dipole vanish is equivalent to the requirement that the difference dipole be a maximum (with respect to unitary transformations within the two-state space).

The assumption that the Platt diabatic states are approximations to the GMH diabatic states should be considered as a physically and chemically motivated assumption. To our knowledge, it has never been established that this condition is a necessary consequence of, or a necessary precondition for, the validity of Brooker's deviation rule.

In order to derive dipole moment descriptors of the bright excitation in our dye set, we point out that the transformation that diagonalizes the dipole moment matrix along a particular direction depends on the ratio of the dipole change to twice the transition dipole in a manner isomorphic to the way the Platt model depends on $\lambda^{Platt}$.[28] To this end, we define a parameter $\lambda^{GMH}$, which specifies transformation that diagonalizes the projected dipole moment matrix as

$$\lambda^{GMH} = \cot 2\theta^{GMH} \equiv \frac{\Delta\mu}{2\mu_{12}} \qquad (10)$$

where $\Delta\mu$ is the difference between projected dipoles in the ground and excited states, and $\mu_{12}$ is the transition dipole between these states. We have also defined the associated mixing angle $\theta^{GMH}$. "GMH" in the superscript refers to the (Cave and Newton's) Generalized Mulliken Hush approach, wherein the relevant diabatic basis is defined so that the diabatic transition dipole vanishes.[28] The approximation of vanishing transition dipole between the diabatic states has been invoked in other two-state models for organic chromophores. For example, Blanchard-Desce and coworkers have written down expressions for dipole elements of push-pull polyenes within this approximation.[19]

We can always divide two real numbers to obtain a third; it is obvious that for any single dye we could write

$$\eta \equiv \frac{\lambda^{GMH}}{\lambda^{Platt}} = \frac{\Delta\mu E_I}{2\mu_{12} b_{LR}} \qquad (11)$$

where $\eta$ is just the ratio of analogous parameters $\lambda^{Platt}$ and $\lambda^{GMH}$. We could then, for a single dye, write the dipole difference and transition dipole by substitution directly into the formulas used in the GMH approach itself.[28] Specifically, for the projected adiabatic difference and transition dipoles we have

$$\frac{\Delta\mu}{M} = \cos 2\theta^{GMH} = \frac{\lambda^{GMH}}{\sqrt{1+\left(\lambda^{GMH}\right)^2}} = \frac{\eta\lambda^{Platt}}{\sqrt{1+\left(\eta\lambda^{Platt}\right)^2}} \quad (12)$$

and

$$\frac{2\mu_{12}}{M} = \sin 2\theta^{GMH} = \frac{1}{\sqrt{1+\left(\lambda^{GMH}\right)^2}} = \frac{1}{\sqrt{1+\left(\eta\lambda^{Platt}\right)^2}} \quad (13)$$

where M is the diabatic difference dipole in the GMH representation.

$$M = \sqrt{(\Delta\mu)^2 + 4(\mu_{12})^2} \quad (14)$$

Platt derived a formula for the transition dipole corresponding to the case $\eta = 1$.[16] This corresponds to identity between diabatic representations in the Platt and GMH models. The two-state model of Blanchard-Desce and coworkers assumes the constraint defining the diabatic states in the GMH model,[7] so the relationships written down by Platt and by Blanchard-Desce and coworkers are the same.

The parameters required to evaluate expressions 11-14 are compiled in Table 2. We have found that equations 12 and 13 do not describe our data set particularly well if $\eta = 1$. In particular, the calculated difference dipoles are approximately half as large as predicted in this limit. There is also a small systematic deviation in the transition dipoles that gets larger as $\lambda^{Platt}$ increases in magnitude.

Equations 12, 13 and 14 should fix the problems associated with non-identity between the Platt and GMH diabatic bases entirely, but at the heavy cost of introducing a new adjustable parameter that varies from dye to dye. This cost would diminish if a functional dependence for $\eta$ on $\lambda^{Platt}$ could be established, which did not itself contain adjustable parameters. We do not pursue this, but instead suggest that, for a set of structurally related dyes, the distribution of $\eta$ over the set the set should be narrow. If so, we can write approximations to the dipole observables in terms of a mean value for the set as

$$\frac{\Delta\mu}{M} \approx \frac{\bar{\eta}\lambda^{Platt}}{\sqrt{1+\left(\bar{\eta}\lambda^{Platt}\right)^2}} \quad (15)$$

and

$$\frac{2\mu_{12}}{M} \approx \frac{1}{\sqrt{1+\left(\bar{\eta}\lambda^{Platt}\right)^2}} \quad (16)$$

where $\bar{\eta}$ is a mean value representing the distribution of $\eta$ over the dye set.

In Figure 4, we show that this approach does afford a good approximation to the dipole moment observables associated with the bright transition for the dyes in our example set. In these plots, the relationships 15 and 16 were fit to the dipole data in our set, using $\bar{\eta}$ as an adjustable parameter. As expected, the best-fit curves are characterized by identical parameters in either case, and are consistent with the observation that the dipole difference is half of the value expected if $\eta = 1$. In this case, the direction along which the quantities are projected is that given by the transition dipole moment itself, because this observable is best defined for dyes near resonance (the projection of the difference dipole in this direction vanishes at resonance).

The physics represented by the parameter $\eta$ appearing in equations 11 in 12 is that of "screening" of the charge-transfer transition via a state-specific redistribution of the electronic density not directly involved in the transfer. This manifests in a state-dependent best effective donor-acceptor distance. It is the same physics that has been discussed by Shin et al. in the context of Ru(III) complexes, where a factor of two difference between experimental difference dipoles and those calculated using a GMH model (using a donor-acceptor distance extracted from crystallography measurements).[36] In our case, state-averaged orbitals are used for both ground and excited states in the SA-CASSCF, and changes consistent with screening of the charge-transfer can be observed in the CI vectors for the states. The argument that this screening should be similar for different molecules in the set ultimately depends on notion that the response of the electronic distribution to the charge transfer is restricted primarily by the topology of the π electron system, which is the same for all molecules we have used in our calculations.[16,19,28,36] For a series of vinylogous dyes, the effect of this screening should decrease as the chain grows longer. This is because the state-dependent variation in donor-acceptor distance is dominated by the polarizability of the heterocycles. The movement of the effective charge centre within the ring will lead to a lower relative error as the length of the molecule increases.

**V. (Hyper)Polarizabilities from the Platt Model**

For organic molecules with a low-lying charge transfer excitation, the (static) polarizability tensor $\alpha$ and (static) first hyperpolarizability tensor $\beta$ have a particularly simple form.[22] In this limit, the tensors are dominated by the vector component along the direction characterizing the charge transfer. The expressions for $\alpha$ and $\beta$ along the direction of charge transfer depend only on the excitation energy, the transition dipole and (for $\beta$) the difference dipole associated with the transition.[11,19,20,22] As we have discussed above, all of these quantities can be expressed or approximately expressed in terms of $\lambda^{Platt}$. We have

$$\alpha = \frac{2\mu_{12}^2}{E_{LR}} = \frac{M^2}{2E_I}\left(\frac{1}{\sqrt{1+\left(\lambda^{Platt}\right)^2}}\right)\left(\frac{1}{1+\left(\eta\lambda^{Platt}\right)^2}\right)$$

$$\approx \frac{M^2}{2E_I}\left(\frac{1}{\sqrt{1+\left(\lambda^{Platt}\right)^2}}\right)\left(\frac{1}{1+\left(\overline{\eta}\lambda^{Platt}\right)^2}\right)$$

(17)

and

$$\beta = \frac{3(\mu_{12})^2 \Delta\mu}{2E_{LR}^2} = \frac{3M^3}{8E_I^2}\left(\frac{\eta\lambda^{Platt}}{\left(1+\left(\eta\lambda^{Platt}\right)^2\right)^{\frac{3}{2}}}\right)\left(\frac{1}{1+\left(\lambda^{Platt}\right)^2}\right)$$

$$\approx \frac{3M^3}{8E_I^2}\left(\frac{\overline{\eta}\lambda^{Platt}}{\left(1+\left(\overline{\eta}\lambda^{Platt}\right)^2\right)^{\frac{3}{2}}}\right)\left(\frac{1}{1+\left(\lambda^{Platt}\right)^2}\right)$$

(18)

where the vector notation subscripts for $\alpha$ and $\beta$ indicate that the expressions are for the component along the direction of charge transfer. In the last part of equations 17 and 18 we have used the "mean $\eta$ approximation" for $\Delta\mu$ and $\mu_{12}$ derived in the previous section.

Notice that each of the expressions in eqns. 17 and 18 have a simple form, which we can write (using $\beta$ as example)

$$\beta = \beta_0 \beta_1(\lambda^{Platt}; \overline{\eta}) \tag{19}$$

where we have separated $b$ into a component $b_0$ that does not depend on $\lambda^{Platt}$

$$\beta_0 = \frac{3M^3}{8E_I^2} \tag{20}$$

and a component $\beta$', which is a function of $\lambda^{Platt}$ and parameterized by $\overline{\eta}$. Note that $\beta_1$ is a dimensionless scalar function, and that it is the same *for any dye in the set*.

This implies that the optimimum magnitude of $\beta$ for any dye occurs at the same value of $\lambda^{Platt}$, and is determined only by the scaling imposed by $\beta_0$.

$$\beta^{opt} \equiv \left|\beta(\lambda^{Platt};\eta)\right|\Big|_{\frac{d\beta}{d\lambda^{Platt}}=0} = |\beta_0|\beta_1^{opt} \qquad (21)$$

Where $\beta_1^{opt}$ only depends on $\overline{\eta}$. Similar conclusions hold for $\alpha$.

Equation 18 predicts that the hyperpolarizability of Brooker dye molecules will increase with increasing chain length. To see this, first note that since $M$ is the donor-acceptor dipole difference in the Generalized Mulliken-Hush model, it should increase roughly linearly in the chain length. Next, note that the isoexcitation *wavelength* of a Brooker dye (for which $b_{LR} = 0$) increases by an approximately constant amount (~ 50 nm) for each additional vinyl group in the chain.[15,16] This behavior is maintained up to some limiting length – estimated to be around 10 methine units[9,37] – where symmetry breaking and soliton formation occur. For ring nuclei in our example set the isoexcitation wavelengths of the monomethine dyes are ~ 450 nm. This implies that the although contribution to $\beta_0$ from the factor $E_I^{-2}$ will have formal contributions with dependence between $N$ and $N^2$, the actual magnitude of this change is will be small for chain lengths less than 10 methine units. Given this reasoning, equation 18 predicts that $\beta_0$ should rise as approximately the third power of the chain length, as is experimentally observed.[5] Note that if $E_I$ decreases while $|b_{LR}|$ remains fixed, then $|\lambda^{Platt}|$ will also be increase. This predicts that the resonance detuning will increase for successive members of a vinylogous series of dyes. Accordingly $\beta_1$ will not stay the same, and will depend on where the vinologous series "starts" – i.e. the value of $\lambda^{Platt}$ for the monomethine (or stilbenoid) member of the series.

Using our expressions for the bond order and bond length alternation, we can also express the static (hyper)polarizability in terms of these quantities, as has also been done in previous studies by Marder and coworkers,[2,8,11] Blanchard-Desce and coworkers,[7,19] and by Lu and Goddard.[20] We can express $\alpha$ and $\beta$ in terms of the bond order alternation coordinate $x$ as

$$\alpha = \frac{2\mu_{12}}{E_{LR}} = \frac{M^2}{2E_I} \frac{1}{\left(1 + \frac{x^2\eta^2}{1-x^2}\right)\sqrt{1 + \frac{x^2}{1-x^2}}} \approx \frac{M^2}{2E_I} \frac{1}{\left(1 + \frac{x^2\bar{\eta}^2}{1-x^2}\right)\sqrt{1 + \frac{x^2}{1-x^2}}} \quad (22)$$

and

$$\beta = \frac{3M^3}{8E_I^2} \frac{x\eta\sqrt{1-x^2}}{\left(1 - \frac{x^2\eta^2}{1-x^2}\right)} \approx \frac{3M^3}{8E_I^2} \frac{x\bar{\eta}\sqrt{1-x^2}}{\left(1 - \frac{x^2\bar{\eta}^2}{1-x^2}\right)} \quad (23)$$

Equations 17-18 and 22-23 are graphed in Figure 5 for $\eta = 0.5$ and 1.0. The overall $x$ dependence is familiar, and has been discussed by Marder and co-workers, by Blanchard-Desce and co-workers, and by Lu and Goddard.[2,8,19,20] The dependency on $\lambda^{Platt}$ has not been discussed before, to our knowledge. In either case the effect of $\eta < 1$ is to concentrate the polarizability $\alpha$ closer to the resonant limit, and the hyperpolarizability $\beta$ farther from it.

The parameters $E_I$ and $M$ chosen for the curves drawn in Fig. 7 are mean values of these parameters over our example set of dyes, so the numerical values of $\alpha$ and $\beta$ shown there are representative of dyes in the set. Equation 18 indicates that the *optimal* magnitude $\beta^{opt}$ increases with increasing $M$ and decreasing $E_I$. The dye in the set with the highest $\beta^{opt}$ is PhO-:PhO-, with $\beta^{opt}(\lambda^{Platt}=0.71;\eta=0.5) = 60 * 10^{-30}$ cm$^5$esu$^{-1}$ and $\beta^{opt}(\lambda^{Platt}=0.50;\eta=1.0) = 87 * 10^{-30}$ cm$^5$esu$^{-1}$. Note that this is the *optimal* value of $\beta$ for this dye – as it symmetric, $\lambda^{Platt} = \beta = 0$ for the *in vacuo* ground state.  These $\beta^{opt}$ place the dyes in this set in the category of molecules with strong nonlinear optical response according to the categorization by Kanis, Marks and Ratner.[5]  They are comparable to short-chain push-pull polyenes, stilbenes and azostilbenes.[1]

**VI. Details of Computations on an Example Dye Set**

In this paper, we analyze relationships between bond alternation, polarity and polarizability that can be derived from the Platt model when additional constraints are assumed to hold for the diabatic states in the model.  We will use quantum chemical calculations on an example set of dyes.  The example set is a complete set of monomethine dyes formed using the heterocyclic nuclei shown in Figure 6.  This set includes several protonation states of the Green Fluorescent Protein (GFP) chromophore motif.[38]  They are good examples because they are derived from only two distinct heavy-atom rings, but they sample a quite large range of $\lambda^{Platt}$.  Optical nonlinearities have been observed in fluorescent proteins (FPs) and their chromophores.[39,40] They exhibit significant solvatochromism,[41] and dipole properties in these systems are amenable to electroabsorption experiments.[42]  Crystallization protocols exist for many FP variants and there are many known crystal structures.[38] The chromophore is covalently bound to the protein in a known orientation; the chromophore frame can be related to the crystal axes frames so that anisotropic electrical properties (e.g. transition dipoles) can be evaluated in straightforward fashion.[43]  The chemical structure of the chromophore as well as its environment can be modified through natural[44] and unnatural[45] amino acid mutagenesis and post-expression synthetic techniques.[46]  In short, these are *superb* systems for the study of structure-property relationships in Brooker dyes.

For each of the dyes generated using the nuclei in Figure 6, we optimized the geometry on the ground state potential surface calculated with MP2 theory[47] and a cc-pvdz basis set,[48] and performed multi-state, multi-reference perturbation theory[49] calculations with a two-state-averaged four-electron, three orbital complete active space self consistent field[50] (SA2-CAS(4,3)) reference state (again with a cc-pvdz basis set). The two states in the average were the ground and first excited states, each of which was given equal weight. Olsen has shown that for the resonant dyes, a three-state average with an identical structure can also be found.[51] From these calculations we extracted second order state energies and first order dipole matrix elements. The bond length data that we present here were extracted from the MP2/cc-pvdz optimized geometries. All calculations were performed using the Molpro software.[52] Data as needed to facilitate the reproduction of calculations, including optimized geometries, SA-CASSCF natural orbitals and occupation numbers are included in a supplement.[53]

Olsen has previously pointed out that the solution of the SA2-CAS(4,3) problem that we use here mirrors the structure of the Platt-Brooker model, because the localized active orbitals are transferrable in the same sense as the Brooker basicity indices.[18] This is illustrated in Figure 7 for a representative asymmetric dye and its parents (a more complete demonstration can be found in figure 4 of reference [18]). For a set of dyes containing the example set used here, a suitably parameterized Platt model reproduces the excitation energies calculated using multi-state multi-reference perturbation theory (on the same SA-CASSCF reference) to within the *a priori* expected accuracy of the calculations themselves.[18] The low-energy excitations of GFP chromophores have been studied multiple times previously using similar computational models.[54,55,56] The authors have previously shown that similar solutions to the analogous three-state problem are an *ab initio* basis for diabatic models of Brooker dye photoisomerization,[57] and that the third state also corresponds to a state predicted by early dye theories.[51] This raises the interesting possibility that structure-property relationships can be established linking geometry, linear and linear optical responses *and non-radiative decay rates*. This is supported by evidence that the proximity to resonance determines the accessibility of alternate twisting pathways.[58]

By comparing our formulas for the dipole observables and polarizabilities against quantum chemical *electronic structure* results, we are ignoring effects of vibronic and solvation coupling. For many systems in the class that we examine, this is a questionable approximation. Painelli and coworkers have emphasized that vibronic and solvation couplings can enhance nonlinear polarizabilities of conjugated push-pull compounds by up to an order of magnitude.[59,60,61] The importance of nonequilibrium solvation has also been highlighted by Hynes and coworkers.[62] We will suggest later that these effects may explain the anomalously high nonlinear polarizability observed for the GFP homologue Dronpa (which is much higher than the maximum possible value suggested by our electronic-only analysis).[39]

## VII. Discussion

We have derived formulas for structure-property relationships within Platt's two-state model[16] of Brooker[15] dyes, and have illustrated these relationships using quantum chemical calculations on a complete set of monomethine dyes related to the GFP chromphore motif. The Platt model provides a natural measure of the deviation from the resonant limit ("cyanine limit") in methine dyes, in the form of the parameter $\lambda^{Platt}$, which for a given dye can be deduced from the linear absorption spectrum of the dye and its symmetric parents. This definition highlights an interesting characteristic of $\lambda^{Platt}$: it is not a property of one dye, but is defined by the properties of different dyes sharing a common set of heterocyclic nuclei. In this sense it is qualitatively different from other descriptors that have arisen in other two-state models of conjugated organic chromophores, such as the bond length alternation or the ratio of transition and difference dipoles. The latter two descriptors are defined in terms of observables estimated on a single molecule.

The relationship between $\lambda^{Platt}$ and the bond length alternation in organic dyes highlights an interesting weakness in the latter descriptor. It shows that the bond *length* alternation relative to a fixed reference characterizes the electronic structure only in an *affine* sense (i.e. without a naturally defined origin). It shows that a better reference state for relating bond length and resonance detuning is that where the bond lengths equal those of the parent symmetric dyes, each on the appropriate domain. This implies that *bond lengths are not necessarily equal at resonance*. The reason is that, if one assumes that the Platt diabatic states represent ideal and complementary single and double bond order alternation, the bond *orders* of a symmetric dye should be exactly one-half.[32,63] However, the bond *lengths* of all possible symmetric dyes are not equal, even on the bridge. Although this point is simple and direct, we have not seen previous explicit discussion of it in the literature. This may be because the observable consequences should diminish for long bridges, which is a common limit to invoke in discussions of the electronic structure of methines. If the bridge is long, the bond lengths near the middle will become independent of the chemical identity of the nuclei at the ends, the reference state for these bonds approaches a standard length alternation.

Our examination of dipole descriptors within the Platt model highlights the relationship between this model and the Generalized Mulliken-Hush approach to the electronic structure of charge-transfer systems.[16,28] In the GMH approach, the diabatic states are defined by a vanishing transition dipole matrix element. Taking a GMH model where the charge-transfer direction is taken along the transition dipole of the dye, it is immediately apparent that the Platt and GMH diabatic states will coincide at the resonant limit, where both the Brooker basicity difference (Platt) and the projected difference dipole (GMH) vanish, yielding $\lambda^{Platt} = \lambda^{GMH} = 0$.

Away from resonance, the diabatic states defined by the Platt and GMH will not generally coincide. We can establish an approximate relationship between the models in the case where the distribution of "screening" parameters $\eta$ is sharply peaked about a mean value $\bar{\eta}$. For the example dye set we use here, the variation of the dipole properties over the set can be described fairly well with $\bar{\eta} \sim 0.5$. This leads to slightly altered expressions for the dipole moment, polarizability ($\alpha$) and first hyperpolarizability ($\beta$) relative to those obtained for $\eta = 1$.[2,19] The expressions we derive for $\eta = 1$ show an identical dependence of $\alpha$ and $\beta$ upon the bond length alternation coordinate as has been presented by Marder and coworkers, Blanchard-Desce and coworkers, Hynes and coworkers and by Goddard and coworkers.[8,19,20,62,64] For $\bar{\eta} < 1$, the expressions we derive lead to higher values of $\alpha$ closer to the resonant limit and move the optimal value of $\beta$ farther away from it.

*The relationships that we have established linking the dipole observables with parameters in the Platt model should be extremely useful in cases where the parent symmetric dyes in a given case cannot be prepared for study.* This will be useful for dyes with difficult syntheses, but is even more important for cases where dyes are coordinated by stereospecific environments, which are not averaged on the timescale of the experiments. A good example of the latter situation is the case of specific binding of a dye to a biomolecule. Biomolecules (and specifically proteins) are

intrinsically non-symmetrical, and in this case it is clear that the parent symmetric dyes will not be easily available. In such cases, our results indicate that the Platt model can be parameterized by experimental determination of the dipole observables (for example, by Stark spectroscopy[65]). Olsen has previously discussed the possibility that the Platt model parameters $b_{LR}$ and $E_I$ may be generalized to describe an ensemble state of a chromophore in condensed matter.[18] Platt himself did this when he showed that the solvatochromism of Brooker dyes could be described by choosing an environment-dependent effective $b_{LR}$.[16]

Boxer and coworkers have measured (by Stark spectroscopy) the length of $|\Delta\mu|$ and the angle spanned by $\Delta\mu$ and $\mu_{12}$ for *A. victoria* GFP and its S65T mutant at 77K, finding in both cases that $|\Delta\mu| \sim 7.0$ Debye and that the angle between $\Delta\mu$ and $\mu_{12}$, $\zeta_A \sim 20^0$ for the "B" absorption band (peaking at 21300cm$^{-1}$ under these conditions). We can use these data to estimate $b_{LR}$ and $E_I$ for a given screening parameter $\eta$ and total dipole magnitude $M$. To see this, note that equations 3 and 12 can be rewritten as

$$\lambda^{Platt} = \sqrt{\left(\frac{E_{LR}}{E_I}\right)^2 + 1} = \frac{\Delta\mu}{\eta M \sqrt{1-\left(\frac{\Delta\mu}{\eta M}\right)^2}} \qquad (24)$$

The B band of GFP is usually assigned to an anionic chromophore,[66] which corresponds to the dye PhO-:ImO- in our example set. Extracting a value of $M$ from the calculation on this dye yields $M = 19.6$ D. The projected value of $\Delta\mu$ onto $\mu_{12}$ taken from the experiment is $\Delta\mu/M = 0.36$. If we know $\eta$ then we know $\lambda^{Platt}$. We can reason that $\eta$ should lie between 0.5 and 1, because if we interpret our calculation as representative of a molecule in ideal isolated vacuum, then the ability of the electrons in the rings to compensate for the charge-transfer transition should be no better than for this case. In a condensed phase environment, this "slop" in the degrees of freedom of the ring electrons should decrease as these degrees of freedom are influenced by interactions in the immediate environment. We then calculate the limits of $\lambda^{Platt}$ under these assumptions as $\lambda^{Platt}(\eta=0.5) = 0.78$ and $\lambda^{Platt}(\eta=1.0) = 0.39$. If we use the experimental absorption peak 21300cm$^{-1}$ as an estimate of $E_{LR}$, we obtain $E_I(\eta=0.5) \sim 16800$cm$^{-1}$, $b_{LR}(\eta=0.5) \sim 13100$cm$^{-1}$ for one end of the range and $E_I(\eta=1.0) \sim 19800$cm$^{-1}$, $b_{LR}(\eta=1.0) \sim 7700$cm$^{-1}$ on the other. In wavelength units, these correspond to isoexcitation wavelengths of 595 and 504 nm, respectively and Brooker deviations of 165 and 34 nm, respectively. These estimates are broadly consistent with known absorption values for proteins carrying a GFP-type chromophore, the reddest absorptions of which peak at ~510nm.[38,67,68]

The static hyperpolarizabilities $\beta$ of two green fluorescent protein variants EGFP[69] and Dronpa have been reported.[39] The distribution of $\beta$ values that we find for our dye set is consistent with the scale of the static hyperpolarizability measured in the experiments on EGFP ($\beta \sim 33 * 10^{-30}$ cm$^5$esu$^{-1}$).[69] The state of the chromophore in EGFP is usually assigned to an anionic state, corresponding to our dye PhO-:ImO-. The measured value of $\beta$ for EGFP is larger than the value we obtain for the dye (PhO-:ImO-) in its *in vacuo* state, but is close to the optimal values of $\beta$ that we estimate for this dye ($\beta^{opt} \sim 33\text{-}48 * 10^{-30}$ cm$^5$esu$^{-1}$ for $0.5 \leq \eta \leq 1$). This suggests that the effective value of $\lambda^{Platt}$ appropriate for the chromophore in EGFP is *not* that of the bare chromophore. It suggests that the chromophore in EGFP has been detuned farther from resonance. A lower effective value of $E_I$ in the protein environment and concurrent elevation of $b_{LR}$ could lead to a higher value of $\lambda^{Platt}$ with only a marginal change in the excitation energy.

Calculated excitation energies of anionic model chromophores (i.e. PhO-:ImO-) are close to the long-wavelength "B" band absorption maxima of GFPs.[66] This has led to the suggestion that the protein has evolved to maintain the chromophore close to its gas-phase electronic structure.[56] However, the same methods[56] predict a dipole difference that is 5-10 times too small to explain electroabsorption results reported by Boxer and coworkers for several GFP variants.[42] Our analysis here suggests that the dispute be resolved by postulating that $b_{LR}$ and $E_I$ are *both* modified by the protein in such a way as to produce a significant change $\lambda^{Platt}$ without effecting a large change in the absorbance wavelength relative to the isolated chromophore. This underscores a simple caveat that should be heeded when using quantum chemistry to assign spectra: *the excitation energy is an insufficient measure of the accuracy of the calculated states*. The incorporation of the protein environment via the embedding in a QM/MM model does *not* lead to a more accurate prediction for the difference dipole associated with the transition.[56] However, immersion of the model chromophore in a polarisable continuum does lead to a difference dipole magnitude that is closer to the electroabsorption result.[70]

The photoproperties of Dronpa protein are more complicated than those of EGFP.[55,71,72] There are two spectroscopic populations in the absorbance spectrum. One of these ("B" band, $\lambda_{max}$~500nm) is commonly assigned to an anionic (PhO-:ImO-) chromophore, while the other ("A" band, $\lambda_{max}$~400nm) is assigned a phenolic neutral (PhOH:ImO-) form.[72] Furthermore, there are *two* distinct populations with "A" band absorbance in Dronpa, which can be accessed by photoswitching or by pH titration, respectively.[72] The static hyperpolarizability of Dronpa is several times larger than the largest calculated hyperpolarizabilities of dyes in our example set, and is different for the two populations with "A" band absorbance.[39] The origin of the enhancement of $\beta$ in Dronpa is currently a mystery. Large nonlinear optical enhancements can occur due to electron-vibration or electron-solvation couplings.[60,61] Dronpa is known to undergo an excited state proton transfer reaction following excitation into the "A" band;[72] no such reaction occurs in EGFP.[71] It is tempting to speculate that strong coupling to proton modes may contribute to the large nonlinear optical response in Dronpa. One observable consequence of such strong coupling would be a noticeable softening and hardening of these modes in the ground and excited state, respectively.[61,73] Quantum interactions with vibrations and solvent modes are not treated by most QM/MM implementations, and so a failure to reproduce the enhanced hyperpolarizability of Dronpa via QM/MM models could be explained via invocation of these effects. Neither the large magnitude of the static hyperpolarizability of Dronpa, nor the difference in hyperpolarizability between photoswitched and pH-adjusted forms was addressed in a series of recent QM/MM simulations of that protein.[55,74]

## VIII. Conclusion

We have derived expressions for bond order alternation, bond length alternation and dipole property descriptors of a methine dye in the context of a two-state model proposed by Platt[16] to describe the color of Brooker dye molecules.[15] We have illustrated and tested these expressions using a quantum chemical data set obtained for a collection of monomethine dyes related to the GFP chromophore motif. We have established a natural origin for bond length deviation coordinates in methine dyes – specifically, the bond lengths of the parent symmetric dyes corresponding to each ring domain. We have clarified the relationships between resonance detuning and dipole properties in monomethine dyes, and have applied these to models of polarizability and hyperpolarizability. We have used the latter to analyze some experimental results on green fluorescent proteins. The relations we have derived may be used as a basis for understanding the behavior of monomethine dyes in related series, so as to guide the design of new dyes or the development of more detailed models.

**Acknowledgement**

This work was supported by the Australian Research Council Discovery Project Program (DP0877875). Computations were done at the National Computational Infrastructure (NCI) Facility, Canberra, with time provided under Merit Allocation Scheme (MAS) Project m03. We thank J.R. Reimers, N.S. Hush and A.N. Tarnovsky for bringing Brooker's work to our attention. We thank A. Painelli, K. Solntsev, L. Tolbert, S. Boxer, W. Domcke, S. Marder, S. Meech, T. Martínez, M. Prescott, M. Robb, M. Olivucci, G. Groenhof, T. Pullerits, T. Smith, M. Smith, S.C. Smith and R. Jansen-Van Vuuren for helpful discussions.

**Table 1a.** Parameters describing the bond length variations in phenoxy nuclei PhO- and PhOH in the context of different monomethine dye pairings

|       | PhO- | | PhOH | |
|-------|------|------|------|------|
| Bond  | $r_i^0/\text{Å}$ | $c_i$ | $r_i^0/\text{Å}$ | $c_i$ |
| C0-C1 | 1.414 | 0.032±0.002 | 1.418 | 0.039±0.003 |
| C1-C2 | 1.441 | 0.017±0.000 | 1.432 | 0.014±0.000 |
| C2-C3 | 1.380 | 0.014±0.001 | 1.388 | 0.012±0.001 |
| C3-C4 | 1.468 | 0.011±0.001 | 1.421 | 0.011±0.001 |
| C4-C5 | 1.466 | 0.012±0.000 | 1.419 | 0.010±0.001 |
| C5-C6 | 1.379 | 0.014±0.001 | 1.385 | 0.012±0.001 |
| C6-C1 | 1.441 | 0.018±0.002 | 1.434 | 0.015±0.004 |
| C4-O4 | 1.252 | 0.017±0.002 | 1.336 | 0.027±0.004 |

**Table 1b.** Parameters describing the bond length variations in imidazoloxy nuclei ImO-, ImOH and ImNH in the context of different monomethine dye pairings.

|       | ImO- | | ImOH | | ImNH | |
|-------|------|------|------|------|------|------|
| Bond  | $r_i^0/\text{Å}$ | $c_i$ | $r_i^0/\text{Å}$ | $c_i$ | $r_i^0/\text{Å}$ | $c_i$ |
| C0-C1 | 1.412 | 0.044±0.003 | 1.372 | 0.043±0.005 | 1.391 | 0.033±0.006 |
| C1-C2 | 1.452 | 0.031±0.003 | 1.477 | 0.020±0.004 | 1.469 | 0.017±0.005 |
| C2-N3 | 1.445 | 0.010±0.002 | 1.452 | 0.008±0.003 | 1.452 | 0.007±0.005 |
| N3-C4 | 1.355 | 0.022±0.005 | 1.342 | 0.012±0.003 | 1.336 | 0.014±0.008 |
| C4-N5 | 1.335 | 0.008±0.004 | 1.325 | 0.017±0.002 | 1.337 | 0.012±0.002 |
| N5-C1 | 1.387 | 0.024±0.003 | 1.405 | 0.016±0.004 | 1.403 | 0.014±0.007 |
| C2-O2 | 1.227 | 0.025±0.003 | 1.209 | 0.019±0.004 | 1.213 | 0.015±0.007 |

**Table 2.** Parameters characterizing the charge resonance/transfer excitation of dyes in the data set, calculated by quantum chemistry. Parameters $b_{LR}$ and $E_I$ of the Platt model are extracted from second-order MS-MRPT2 state energies. Difference dipole norms $|\Delta\mu|$, transition dipole norms $|\mu_{12}|$ and the angle subtended by the difference and transition dipoles $\zeta_A$ are evaluated from first-order MS-MRPT2 state properties. Energies were extracted from second-order MS-MRPT2 energies.

| Left Nucleus | Right Nucleus | $b_{LR}$ (cm$^{-1}$) | $E_I$ (cm$^{-1}$) | $|\Delta\mu|$ (D) | $|\mu_{12}|$ (D) | $\zeta_A$ (deg) |
|---|---|---|---|---|---|---|
| PhO- | PhO- | 0 | 18224 | 0.2 | 11.0 | 90.0 |
| PhO- | PhOH | 22618 | 19173 | 8.8 | 7.8 | 4.4 |
| PhO- | ImO- | 1142 | 20462 | 1.7 | 9.7 | 11.7 |
| PhO- | ImOH | 20258 | 21048 | 6.9 | 8.6 | 2.4 |
| PhO- | ImNH | 8384 | 18839 | 2.7 | 9.4 | 30.5 |
| PhOH | PhOH | 0 | 20226 | 0.3 | 9.5 | 89.9 |
| PhOH | ImO- | 19907 | 21666 | 5.7 | 7.4 | 0.1 |
| PhOH | ImOH | 1822 | 22324 | 1.1 | 9.1 | 12.5 |
| PhOH | ImNH | 11969 | 19855 | 7.0 | 8.3 | 5.4 |
| ImO- | ImO- | 0 | 23327 | 0.3 | 8.4 | 90.0 |
| ImO- | ImOH | 18288 | 24091 | 5.7 | 7.0 | 4.1 |
| ImO- | ImNH | 5223 | 21241 | 2.6 | 7.8 | 25.4 |
| ImOH | ImOH | 0 | 24907 | 0.7 | 8.3 | 89.9 |
| ImOH | ImNH | 13658 | 21873 | 5.8 | 7.3 | 1.6 |
| ImNH | ImNH | 0 | 19498 | 1.2 | 8.3 | 90.0 |

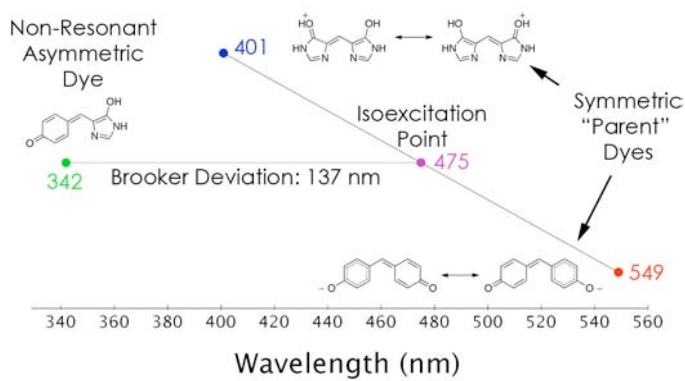

**Figure 1**. "Brooker's Deviation Rule" is a key concept in the resonance color theory of dyes. The empirical rule states that the absorbance maximum of an asymmetric dye is no redder than the wavelength at its "isoexcitation point", which is the mean wavelength of its symmetric "parent" dyes. In energy units, the corresponding concepts are the "Brooker basicity difference" $b_{LR}$ and the "isoexcitation energy" $E_I$ (c.f. eqn. 3). A dye whose absorption is equal to (or close to) its isoexcitation point is "resonant". The deviation of the absorbance of a dye relative to its isoexcitation point is called the "Brooker Deviation", and is a measure of the detuning from resonance. The Brooker deviation is a measure of the difference in "basicity" of the rings, and can be correlated with other chemical measures for electron withdrawal and donation capacities such as the Hammet scale. In this paper, we establish relationships between the Brooker basicity and bond alternation measures, oscillator strengths and non-linear polarizabilities for Brooker dyes.

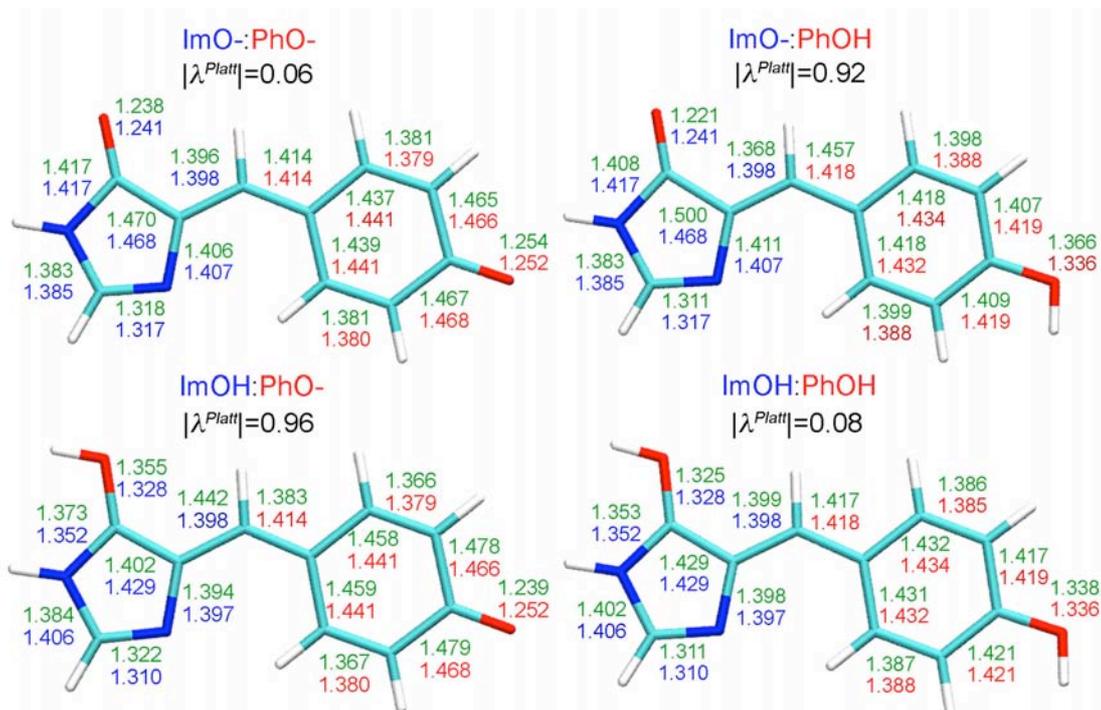

**Figure 2.** The bond lengths of a resonant asymmetric dye converge to those of its symmetric dyes on the corresponding ring domains. Here, we show two resonant symmetrically protonated states of a phenoxy-imidazoloxy dye (ImO-:PhO-, top left and ImOH:PhOH, bottom right), and two non-resonant asymmetrically protonated states with opposing bond alternation (ImOH:PhO-, bottom left and ImO-:PhOH, top right). Bondlengths of the asymmetric dye are shown in green and those of its bis-imidazoloxy and bis-phenoxy parent dyes are shown in blue and red, respectively. For the resonant dyes, the bondlengths on each domain are no more than 0.003Å different from the respective parent dyes. Deviations of the non-resonant dyes from the symmetric parents are generally an order of magnitude greater. Geometries were optimized using MP2 theory and a cc-pvdz basis set (c.f. Section VI). We stress that the bond lengths of an asymmetric resonant dye *are not equal*, even on the bridge. The appropriate coordinate to describe detuning from resonance is not a strict bond alternation, but an alternating deviation from the bond lengths of the parent dyes.

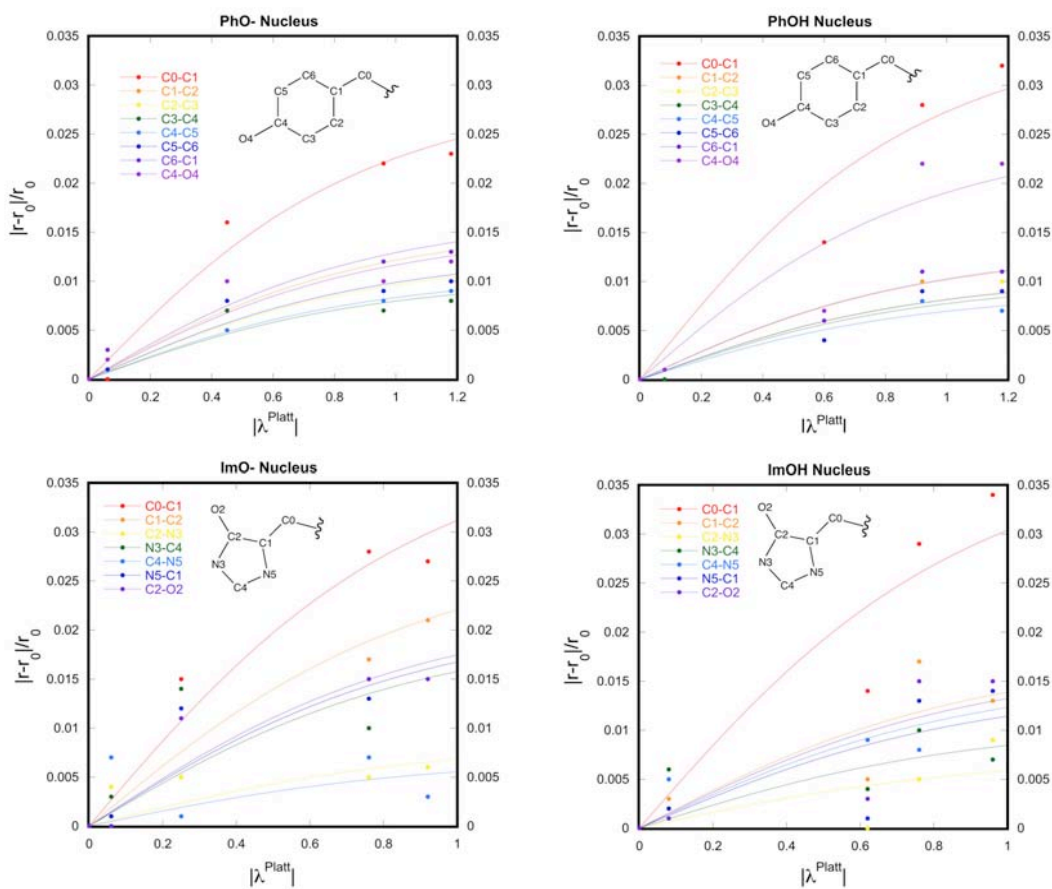

**Figure 3.** Detuning from resonance produces an alternating deviation in the bondlengths relative to the symmetric parent dyes. Deviation absolute magnitudes for four different nuclei are plotted against the absolute value of the detuning parameter, determined by the conjugate nucleus in a monomethine dye. Data obtained from optimized geometries are fitted to a function linear in the bond order alternation parameter $x$ (eqn. 7). Fitted scaling constants and errors are listed in Tables 1a and 1b. Data were generated as described in Section VI.

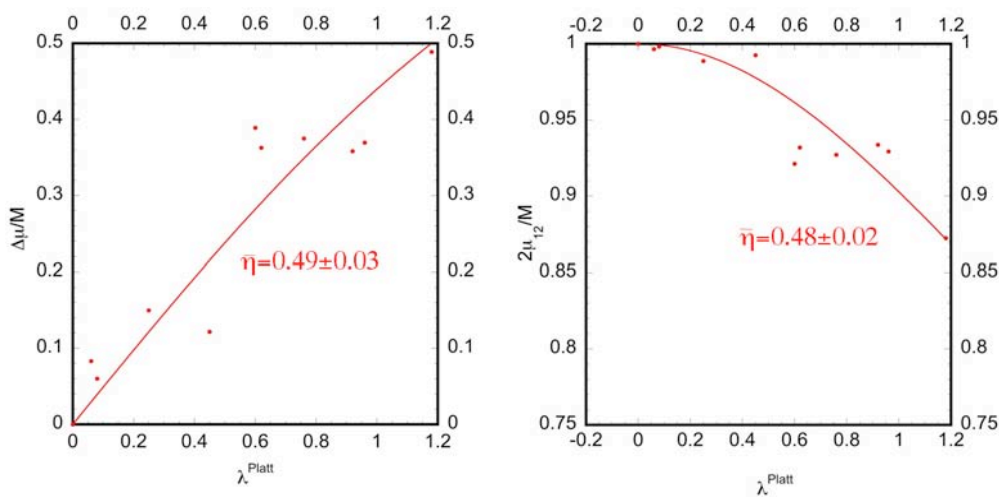

**Figure 4.** Projections of the adiabatic $S_1$-$S_0$ difference dipoles (left) and transition dipoles (right) along the direction of the transition dipole. Data are shown alongside best one-parameter fit plots of the functional form of the dipoles derived from the Platt model (eqns. 15 and 16). The figure demonstrates that dipole properties for different dyes in the set can be reasonably approximated by a single structure-property relationship using an effective screening parameter $\bar{\eta}$ (eqn. 11) extracted from the entire set. Data were generated as described in Section VI.

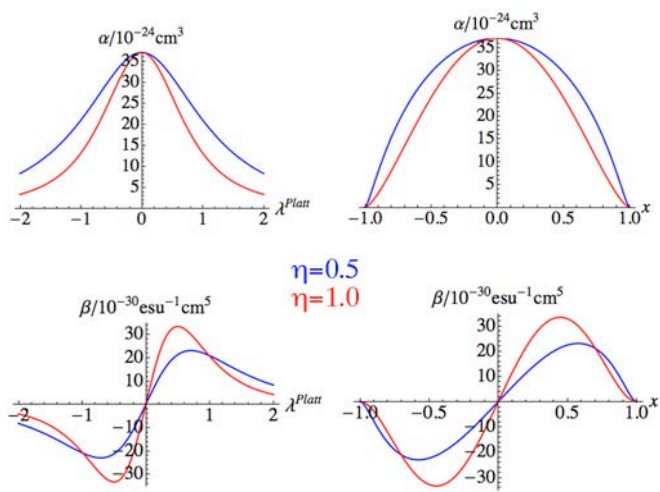

**Figure 5**.  First (a, top) and second (b, bottom) order polarizabilities as a function of the Platt mixing parameter $\lambda^{Platt}$ (left) and the bond order alternation coordinate $x$ (right), for different values of the screening parameter $\eta$, which is a measure of the difference between diabatic states defined by the Platt and Generalized Mulliken Hush models (eqn. 11).  The set of dyes used as example here fall in a narrow range around $\eta$=0.5 (blue line).  The case of $\eta$=1.0 (red line) corresponds to identity between diabatic states in the Platt and Generalized Mulliken-Hush models. Decreasing the value of $\eta$ causes $\alpha$ to peak more broadly, increasing its magnitude for dyes farther from resonance.  Likewise, the extrema of $\beta$ move farther from the resonant limit.  Note that $\beta$ vanishes for a resonant dye, independent of $\eta$.  Units are those appropriate for the electrostatic (esu) CGS unit system.  They are often abbreviated as just "esu" in the literature (for polarizabilities of any order). Note that the scale of $\alpha$ and $\beta$ are set by the parameters $E_I$ (isoexcitation energy) and $M$ (Generalized Mulliken-Hush diabatic difference dipole) (c.f. eqns. 17 and 18).  For this figure, we set $E_I = 21100 cm^{-1}$ and $M = 6.9D$, which are mean values of these parameters over the example set of dyes.  The scale of $\alpha$ and $\beta$ shown in the figure is therefore representative of our example dye set.

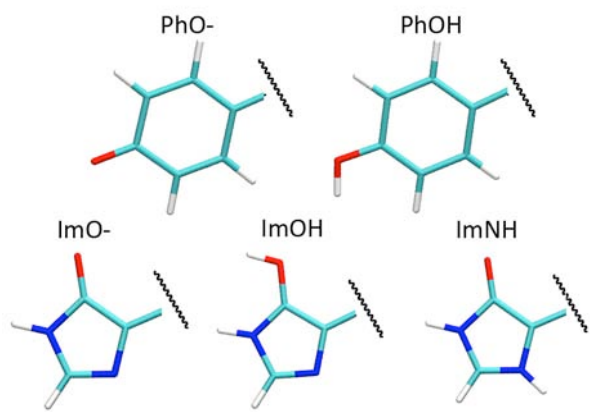

**Figure 6.** Our example data set consists of calculations on a complete set of monomethine dyes generated by pairing five distinct heterocyclic nuclei, shown here. The set thus generated includes 15 chemically distinct dye structures. We refer to the dyes using the notation "L:R" where L and R are the left and right nuclei, respectively.

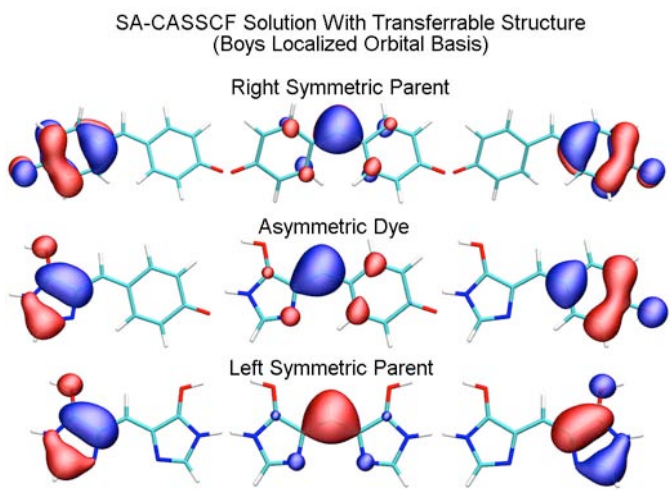

**Figure 7.** For each dye in the data set, excitation energies and state dipole properties were generated with multireference perturbation theory calculations using a two-state averaged complete active space self-consistent field (SA-CASSCF) reference space with four electrons in three orbitals. The SA-CASSCF solution family has been discussed previously; its orbital structure is analogous to the conceptual structure of resonance color theory. The Boys-Localized active space orbitals on each ring have an approximately transferrable structure over the different dyes in the set, and target analogous valence states for each dye. Active space orbitals (both natural and localized) for all dyes in the set are shown in the Supplement [ref. 53].